\documentclass{svjour2}                   
\smartqed  
\usepackage{amsmath,graphicx,color}

\begin{document}

\title{Longitudinal inverted  compressibility in super-strained metamaterials 
}

\titlerunning{Longitudinal inverted  compressibility}

\author{Zachary~G.~Nicolaou   \and  Adilson~E.~Motter}

\authorrunning{Zachary G. Nicolaou   \and  Adilson E. Motter}

\institute{Z. G. Nicolaou\at
              Department of Physics, California Institute of Technology, Pasadena, CA 91125, USA \\
              \email{znicolao@caltech.edu}         
           \and
           A. E. Motter \at
           Department of Physics and Astronomy and  Northwestern Institute on Complex Systems, Northwestern University, Evanston, IL 60208, USA\\
              Tel.: +1-847-491-4611\\
              \email{motter@northwestern.edu}   
}

\date{Received: date / Accepted: date}

\maketitle

\begin{abstract}
We develop a statistical physics theory for  
solid-solid phase transitions in which a 
metamaterial 
undergoes longitudinal  contraction 
in response to increase in external tension.
Such transitions, which are forbidden 
in thermodynamic equilibrium, 
have recently been shown to be possible  
during the decay of metastable, super-strained states.  We present a first-principles model to predict these transitions and validate it using molecular dynamics simulations. 
Aside from its immediate mechanical implications,  
our theory  
points to a wealth of analogous inverted responses, such as inverted susceptibility or heat-capacity transitions, allowed 
when considering realistic scales.

\end{abstract}

\section{Introduction}
\label{intro}

\begin{figure}[t!]
\begin{center}
\includegraphics[width=9cm]{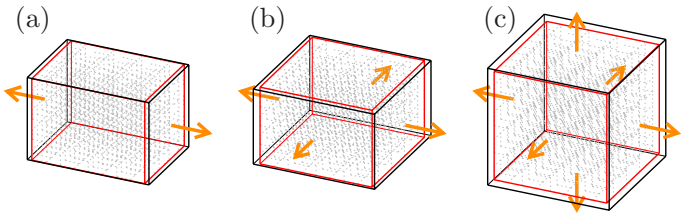}
\end{center}
\caption{Inverted compressibility transitions in a 3D  material in response to  (a)~uniaxial, (b) biaxial, and (c) isotropic tensional stresses (arrows).
 The external solid lines indicate the boundaries of the super-strained material prior to the increase in stress and the internal solid  lines indicate the post-transitional state.  In each case, the material contracts along the same directions as the increasing applied tension during the transition. The material and simulations are defined in the text. 
\label{fig1}
}
\end{figure}

For a closed thermodynamic system in equilibrium, the volumetric compressibility $\kappa=-\frac{1}{V}\frac{\partial V}{\partial P}$ is strictly non-negative because of stability constraints.   Such constraints are manifest through 
first-order phase transitions 
and phase separation via the Maxwell construction that effectively avoids  any branch in the $PV$ diagram that would correspond to negative $\kappa$.  
Furthermore, during any equilibrium phase transition, the occupied volume always decreases as pressure increases and vice versa \cite{landau}.  This analysis  
implicitly assumes that stress changes infinitely slowly. However, when the stress changes at a finite rate, it is common for a solid \cite{salje} (and also for polymers \cite{Schnurr2002}) to occupy a metastable configuration for an extended period before undergoing a phase change, and hence the previous stability constraints no longer apply. We 
posit 
that, if an increase in tension causes the previous equilibrium state to become metastable,  the solid 
may undergo a transition in which the decay of this super-strained state results in a decrease in volume as the tension increases. 
We refer to this process as an inverted compressibility transition (ICT).  
The existence of such transitions has been numerically demonstrated in Ref.~\cite{natmat2012} for 1D and 2D systems. In this paper,  we 
 model from first principles  
a related class of metamaterials that exhibit ICTs in any dimension,  as simulated in Fig.~\ref{fig1}.

\section{Model System}
\label{sec2}

\begin{figure}[t!]
\begin{center}
\includegraphics[width=8cm]{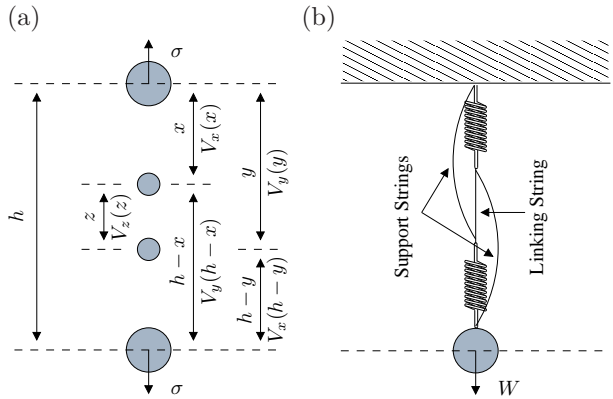}
\end{center}
\caption{(a) Constituent of materials that undergo ICTs. The inter-particle distances and potentials are indicated on the left and right
of the arrows, respectively. (b) Spring-string system in which the weight $W$ 
 is hung by two identical springs, one linking string, and two
identical supporting strings that initially carry no tension.
The weight may rise when the linking string is cut.
\label{fig2}}
\end{figure}

To develop our theory, 
 we consider a crystal defined on a regular lattice in $D$ dimensions.  The constituents are shown in Fig.~\ref{fig2}(a). They consist of four collinear particles interacting through a potential energy $V(x,y,h; \sigma)=V_x(x)+V_y(y)+V_z(y-x) +V_x(h-y)+V_y(h-x) - \sigma h$, where $x$, $y$, and $h$ are inter-particle separations, $V_x$, $V_y$,  and $V_z$ are pairwise interaction potentials, and $\sigma$ is the applied tensional stress.  The two external particles correspond to lattice points, while the two internal ones correspond to edges of the lattice. We consider potentials such that, depending on $\sigma$ in the lattice, the constituents may occupy one of two configurations provided the temperature $T$ is not too high. When $\sigma$ is small, a weak $V_z$-bond forms between the internal particles of each constituent, resulting in a {\it coupled state};
 when  $\sigma$ is increased,
 this bond dissociates through a decoupling transition, resulting in a {\it decoupled state}.
This can be easily achieved with nonlinear interactions. We note that nonlinear interactions are receiving increasing attention in the context of mechanical materials---e.g., in the study  of  solitons \cite{Nesterenko}, particularly in applications that manipulate acoustic response \cite{Spadonia2010,Boechler2011}, and in the study  of   biomechanical networks \cite{Sheinman2012,Broedersz2008}.

During a decoupling transition of the system in Fig.~\ref{fig2}(a), an increase in the applied stress causes a switch from one equilibrium 
$(x,y,h)=(x^*, y^*, x^*+y^*)$ to a different equilibrium, $(\tilde{x}^*, \tilde{y}^*, \tilde{x}^*+\tilde{y}^*)$. Since the derivative of the potential 
energy $V(x,y,h; \sigma)$  with respect to $h$ vanishes at both equilibria, it follows that 
\begin{equation}
V_x{'}(x^*)+V_y{'}(y^*)=V_x{'}(\tilde{x}^*)+V_y{'}(\tilde{y}^*).
\end{equation}
 If $V_x{'}$ and $V_y{'}$   
are monotonically increasing (i.e., spring-like) we can conclude that  $\tilde{x}^*\leq x^*$ and $ \tilde{y}^* \geq y^*$ (or vice versa). Then, if 
the potentials are chosen such that the decrease in $x$ overcomes the increase in $y$, the transition will indeed be an ICT.
This will follow when $V_y$ hardens just after the transition (like a string becoming taut), and the effect can be 
optimized further if $V_x$ hardens before the transition. Specific choices for the potentials used in our numerics and that result in ICTs are presented in  Sec.~\ref{modelval}. 

\begin{figure}[t!]
\begin{center}
\includegraphics[width=11cm]{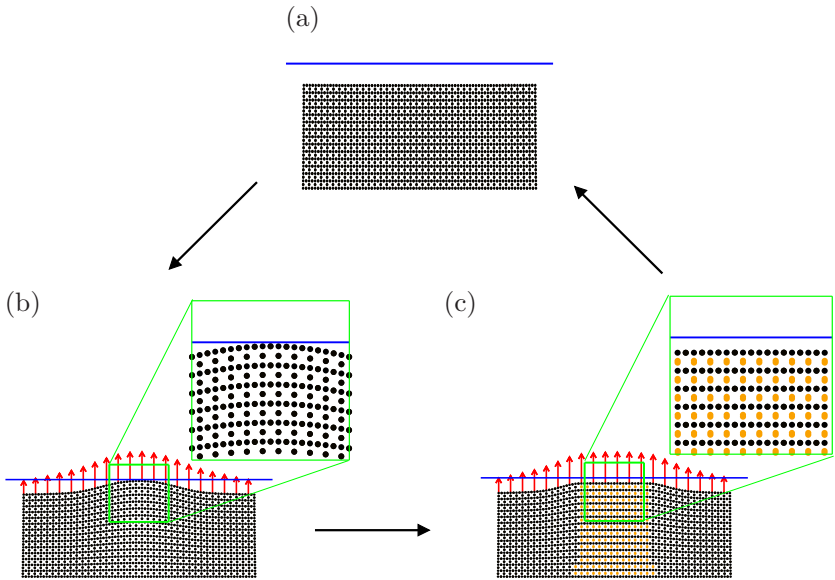}
\end{center}
\caption{
(a) A two-dimensional  material consisting of a square lattice of the constituents in Fig.~\ref{fig2}(a) 
for zero applied stress. 
Same system (b) just prior to  and (c) just after an ICT, 
where the red arrows indicate the applied force profile 
and black (orange) dots indicate coupled (decoupled) states.
The material touches the blue reference line in (b) but is entirely below it in (c), even though the applied tension was increased.
The  insets are magnified to illustrate this. 
When the applied tension is relaxed back to zero, the system returns to its initial, rectangular configuration (Color figure online). 
\label{fig8}
}
\end{figure}

The decoupling transition
that occurs when $\sigma$  is increased can be interpreted  more concretely in terms of the spring-string system
introduced by Joel E. Cohen and Paul Horowitz  \cite{Cohen91} and reproduced in
 Fig.~\ref{fig2}(b): when the linking string is cut, the initially slack support strings become taut and the mass can, counterintuitively, rise. 
This occurs because the springs go from a series configuration to a parallel configuration, causing the two springs to contract as the load becomes distributed between them.  Initially each spring must hold the entire weight $W$,  leading to an equilibrium height  $h=l_0 + l-\delta+\frac{W}{k}$, where $k$ is the spring constant,  $l_0$ is the length of the unstretched springs,  $l$ is the length of the support strings, and $\delta$ is the amount of slack the strings have.  Once the linking string is cut, each spring must hold only half of the weight $W$, which leads to an equilibrium height $h=l_0 + l+\frac{W}{2k}$.  Thus the mass rises by $\Delta{}h=\frac{W}{2k}-\delta$ when the linking string is removed, which is positive provided that $\delta$ is sufficiently small. This behavior has long been known \cite{Cohen91}. 
Nevertheless,  likely owning to the fact that it involves an irreversible change of the system, 
 this process has not been widely 
considered as a mechanism to design materials exhibiting  
mechanical behavior not  found in natural materials. The ICTs undergone by the system in Fig.~\ref{fig2}(a), on the other hand, only involve reversible changes and can be repeated by only varying the external force $\sigma$, as illustrated in Fig.~\ref{fig8} for a two-dimensional lattice of such constituent systems.

\section{Equilibrium Properties}

The equilibrium properties of the proposed materials can be derived from the Helmholtz free energy, $A=U-TS$, where $U$ is the internal energy and $S$ is the entropy. To determine $A$ given some strain $\epsilon$, taken as deformation relative to zero-stress linear size,
we first analyze the low temperature limit. 
For the materials we consider, ignoring any zero-point kinetic energy,  to lowest order
the free energy equals the sum of the potential energies of the constituents, $A=V_{tot}$.  The thermodynamic equilibrium is then
the state in which every constituent occupies the lowest-energy mechanical equilibrium that conforms to the external constraints. Any other mechanical equilibria are local minima of the free energy, corresponding to metastable states. 
Given that only the coupled equilibrium exists for small stresses and only the decoupled equilibrium exists at large stresses, it follows that the coupled and decoupled equilibria have equal potential energy at some intermediate stress. Equilibrium statistical mechanics predicts that a phase transition will occur at this point as either the stress or the strain is varied. However, ignoring zero-point quantum fluctuations, the lifetimes of
metastable states
go to infinity as temperature goes to zero.

Further insight into this problem is obtained by considering the Debye model \cite{landau}, which is rigorously correct in the low temperature limit and allows us to explicitly account for the effect of temperature. 
It is convenient to initially assume that all constituents are in either the coupled or decoupled state.
The constant-strain specific heat capacity of a crystal is asymptotically $c_\epsilon=\alpha{}k_B\left(\frac{T}{\Theta}\right)^D$, where $k_B$ is the Boltzmann constant, $\Theta$ is the Debye temperature, proportional to the speed of sound waves in the system, and $\alpha$ is a numerical constant. 
Using this to determine $S$ and $U$, we find that $A=V_{tot}-\frac{\alpha}{D(D+1)}\left(\frac{T}{\Theta}\right)^D(k_BT)$, which now accounts for the internal energy and the entropy of the system at nonzero temperature. This analysis breaks down only near spinodal points (i.e., points at which the coupled or  decoupled metastable states cease to exist), where any deviation away from $T=0$ will lead to a decay of the metastable states. At all other points $\Theta \neq{}0$, and hence  the free energy is well-approximated by just the potential energy for sufficiently small $T$.  This implies that the local minima of the free energy for small $T>0$ will correspond to those of the potential energy, so that the metastable states of pure phase indeed persist for small $T>0$.   

For states of mixed phases, with proportion $\chi$ in the coupled state and proportion $1-\chi$ in the decoupled state, the free energy contains contributions from each of the separate phases plus an additional contribution due to the entropy of mixing, 
\begin{align}
A= & \mbox{$\chi \left[V_{tot,c} -\frac{\alpha}{D(D+1)}\left(\frac{T}{\Theta_c}\right)^D(k_BT)\right]$}\nonumber\\ 
& +  \mbox{$(1-\chi) \left[V_{tot,d} -\frac{\alpha}{D(D+1)}\left(\frac{T}{\Theta_d}\right)^D(k_BT)\right]$}\nonumber \\
& +   \mbox{$k_B T\left(\chi \log \chi + (1-\chi) \log (1-\chi)\right)$}, 
\end{align}
where the subscripts $c$ and $d$ indicate the values in the coupled and decoupled states, respectively.  The thermodynamic equilibrium is the state that minimizes $A$ with respect to $\chi$, which for small $T$ is $\chi = \frac{\exp (k_B (V_{tot,d} - V_{tot,c})/T)}{1+\exp (k_B (V_{tot,d} - V_{tot,c})/T)}$ to leading order. Since $\chi$ will now vary continuously from nearly $1$ to nearly $0$  
as the stress is increased, there is technically no equilibrium phase transition for any $T>0$.  
 However, if this change
 occurs over a scale $\Delta \sigma$ which is below experimental resolution, the behavior will be indistinguishable from a first order phase transition. 
Furthermore, as noted before metastability persists for small $T>0$, so that when the applied stress or strain is varied at a finite rate, the response is hysteretic, just as in an ordinary first-order phase transition.
We will assume that $T$ is sufficiently small that this is the case and will abuse language by referring to the transition as a ``phase transition'' for $T>0$ throughout the paper.

\section{Inverted Compressibility Transitions}
\label{sec3}

\begin{figure}[t!]
\begin{center}
\includegraphics[width=8cm]{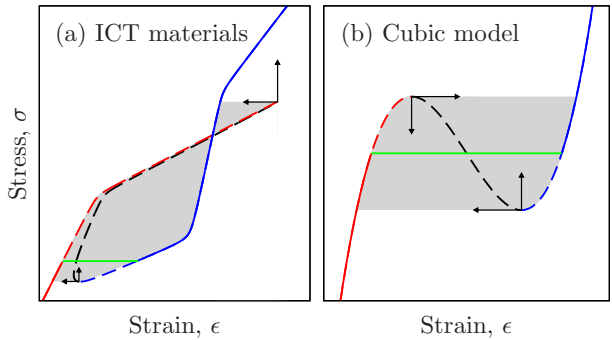} 
\end{center}
\caption{Hysteresis loops for (a) 
materials exhibiting ICTs
and (b) the cubic model.  Solid lines are the thermodynamic equilibria, 
color dashed lines are metastable pure states, and black dashed lines are unstable states.  The gray area includes all possible metastable phase mixtures, and arrows indicate transitions at spinodal points (horizontal when stress is controlled and vertical when strain is controlled) (Color figure online). 
 \label{fig3}}
\end{figure}

Figure \ref{fig3} contrasts the hysteresis loop for materials  that undergo ICTs (of which the materials considered here are a specific case)  with the commonly used cubic model for solid-solid phase transitions, which does not exhibit ICTs. Following the figure, suppose the system begins in the lower left equilibrium (red solid line) as the stress is increased from small values.  
As the stress increases, it becomes possible to occupy a ``flag'' of metastable mixed states, and the system will drift into the gray area at a rate that depends on 
temperature and sample size, as quantified below.
When the temperature is sufficiently small, the state of the system remains very near the left pure metastable state (red dashed line) until it reaches the vicinity of the spinodal point at the top of the gray area.  At this point, the metastable state must decay, as indicated by the arrows. For the proposed materials [Fig.~\ref{fig3}(a)], this transition is an ICT, so that the strain decreases as the stress increases (top horizontal arrow), while for the cubic model [Fig.~\ref{fig3}(b)], the strain increases as the stress increases.  When it is the strain that is controlled instead, it is also easy to deduce a force amplification phenomenon in which increase in strain leads to a transition to large restoration stress in these materials [top vertical arrow in Fig.~\ref{fig3}(a)], which is not present in the cubic model or conventional materials [top vertical arrow in Fig.~\ref{fig3}(b)].

We verify the above conclusions quantitatively by analyzing the decay of metastable states, following the general approach in solid-solid phase transitions of modeling nucleation by thermal activation \cite{abeyaratne}.  We focus on a cell of our material's $D$-dimensional lattice that has  $N$ constituents. 
For simplicity, we consider the case of forces applied equally along all $D$ dimensions of the lattice.
The most important macroscopic timescale is the volume fluctuation rate $\tau_V$, and  the temperature scale above which metastable states are generally short-lived is $T_0$ (see Appendix). We non-dimensionalize times and temperatures by expressing them in units of these references. We note that for individual constituents,  the relevant timescales are much shorter  than $\tau_V$.

\section{Statistical Physics Model}

\subsection{Microscopic Dynamics}

\begin{figure}[t!]
\begin{center}
\includegraphics{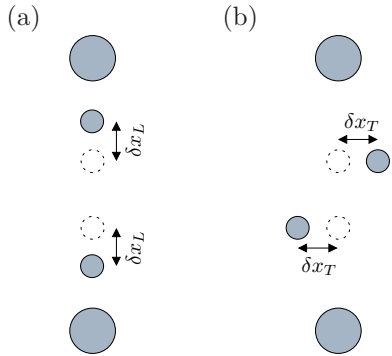}
\end{center}
\caption{
Modes of oscillation leading to  coupling and decoupling transitions:  (a) the longitudinal mode and (b) the $D-1$ transverse modes 
in which the central particles of the constituents move in opposite directions.
 \label{fig6}}
\end{figure}

To model the transitions between coupled and decoupled states, we focus on an individual constituent near a mechanical equilibrium 
$(x^*,y^*, h^*)$, and treat the remainder of the lattice as a heat bath in thermal equilibrium with this constituent. The constituent is assumed to have a  fixed length $h=h^*$, which corresponds to a canonical ensemble for its interaction with the heat bath.
 Figure \ref{fig6} illustrates the vibrational modes of such a constituent  in which the central particles move in opposite directions.
 These modes are of particular interest because they are precisely the vibrations that give rise to the coupling and decoupling transitions of the constituents, and thus the frequencies of these modes indicate how often such events may occur. 
 
To calculate the frequencies of these modes, we  linearize the equations of motion assuming that the displacements from equilibrium $\delta x_L$ and $\delta x_T$ are small (Fig.~\ref{fig6}). We find that the respective angular frequencies of the longitudinal mode and transverse modes are
 \begin{eqnarray}
 \label{long}
&&\omega_L=\sqrt{\frac{k_x+k_y+2k_z}{m}}, \\
&&\omega_T=\sqrt{\frac{\frac{F_x}{x^*}+\frac{F_y}{y^*}+2\frac{F_z}{y^*-x^*}}{m}},
\label{long2}
\end{eqnarray}
where $m$ is the (assumed to be common) mass of the constituent particles; the $F$'s are the first derivatives and the $k$'s are the second derivatives of the corresponding potentials evaluated at the mechanical equilibrium point. We note that from the viewpoint of the entire lattice, such oscillations appear as optical modes of vibration.
By focusing on individual constituents we have effectively restricted attention to optical vibrations with zero wavenumber to estimate  the microscopic timescales.

We assume that the timescale for which stress or strain is varied is much larger than $\tau_V$. Under this condition, the quasi-static approximations are valid and the material can be assumed to be always near (metastable) equilibrium, except during the transition events. In order for a constituent to actually transition between the coupled and decoupled states, thermal fluctuations must be sufficiently large to overcome the free energy barrier separating the metastable state from the thermodynamic equilibrium state.  We approximate this free energy barrier as twice the potential energy differences between the intermediate unstable equilibrium [black dashed line in Fig.~\ref{fig3}(a)] and the coupled ($2V_b^c$) or decoupled ($2V_b^d$) stable equilibrium of the constituents (assuming they are initially coupled or decoupled, respectively). 
The factor two arises because the kinetic and potential energy contribute equally to the total energy, so only $\frac{k_BT}{2}$ of the energy is typically distributed as potential energy.  
This
slightly subtle
 factor of two was verified as an intermediate step in our numerical simulations below.

\subsection{Rate of Transition Events}

We now  calculate the expected rate of events that overcome the free energy barrier. The probability that thermal fluctuations a constituent exceed an energy of $2V_b$ is $\exp[{-\frac{2V_b}{k_BT}}]$, as  determined by the canonical ensemble. The frequency at which the energy barrier is approached is derived from the frequencies in Eqs.~(\ref{long})-(\ref{long2}). 
During a decoupling transition, both transverse and longitudinal vibrations strain the $V_z$ bond, so we take the effective rate of energy barrier approach as $\left[\omega_L+(D-1)\omega_T\right]$.  For a coupling transition, we must have a coordinated approach in both the longitudinal and transverse directions  in order for the central particles to come together, so we use a reduced frequency $\left[\omega_L^{-1}+(D-1)\omega_T^{-1}\right]^{-1}$.  The rate of decoupling transitions for each constituent is then  
\begin{equation}
f^d(\sigma,T) = 
\frac{\omega_L(\sigma)+(D-1)\omega_T(\sigma)}{2\pi}\exp\left[{-\frac{2V_b^c(\sigma)}{k_BT}}\right], 
\end{equation}
and the rate of coupling transitions is 
\begin{equation}
f^c(\sigma,T) = \frac{\omega_L(\sigma)\omega_T(\sigma)}{2\pi[(D-1)\omega_L(\sigma)+\omega_T(\sigma)]}\exp\left[{-\frac{2V_b^d(\sigma)}{k_BT}}\right].
\end{equation}

To proceed,
suppose we increase the applied tensional stress on the $N$-constituent cell  from $\sigma=0$ at a constant rate $\nu_{\sigma}=d\sigma/dt$. All of the constituents in the cell are initially coupled. However, decouplings become possible when the stress increases past some $\sigma^c$, where the decoupled equilibrium appears, and necessarily occur by some $\sigma^d$, where the coupled equilibrium ceases to exist.
Then, applying our estimate for the transition rate for each constituent, the expected number
of decoupling transitions that will occur by the time the stress reaches some $\sigma$ is given by the integral of the rate over time
\begin{equation}
n^d(\sigma ;N,\nu_\sigma,T)=N\int_{0}^t{f^d(\sigma(t'),T)dt'}=N\int_{\sigma^c}^\sigma{f^d(\sigma',T)\frac{d\sigma'}{\nu_\sigma}},
\label{decouple}
\end{equation}
where $\sigma^c<\sigma  <\sigma^d$. 
 In the second equality, we have changed the integration variable from time to stress and used the fact that $dt'=d\sigma'/\nu_\sigma$.  
Equation (\ref{decouple})  should be interpreted as a valid estimate of the first transition event,
which is all we need to complete our model for the decay of metastable states. Specifically, we note that the transition events are largely irreversible 
(they tend to go only from metastable to stable states) and different transition events are not independent in the material: a single coupling or decoupling transition produces a cascade of further transitions, and macroscopic metastability can be completely and suddenly lost. Therefore, we take Eq.~(\ref{decouple}) as an estimate for the beginning of the decay of metastability. The time evolution of the phase transition would require  the description of a kinetic relation \cite{abeyaratne,Krapivsky2010}, but such details are not necessary here.  So long as $n^d(\sigma; \cdot)\ll{}1$, we can assume that no decoupling events will occur in the cell as the system is stressed from $\sigma^c$~to~$\sigma$. 

Taking all this together, we model the stress at which metastability is lost, $\sigma^*$, through the implicit equation
\begin{equation}
n^d(\sigma^* ;N,\nu_{\sigma},T)=1,
\label{model}
\end{equation}
which we use to quantify the ICT effect.  By replacing  $f^d$ with  $f^c$  and/or $\sigma^*$ with  $\epsilon^*$ 
and appropriately changing the limits of integration in the model above, similar equations can be derived for coupling transition and also for strain-controlled processes.

\section{Model Validation and Application}
\label{modelval}

We have verified that our model accurately predicts the onset of the decoupling transitions, and ICTs in particular, using  molecular dynamics (MD) simulations.
The MD simulations were carried out using Nos\'{e}-Hoover chains \cite{Martyna92}  and a generalization of the Andersen barostat \cite{Andersen80}. 
Incidentally, Figs.~\ref{fig1} and~\ref{fig8} were generated using similar MD simulations.
For details on our implementation of the MD simulations, see Ref.~\cite{natmat2012}. 

\begin{figure}[t!]
\begin{center}
\includegraphics[width=10cm]{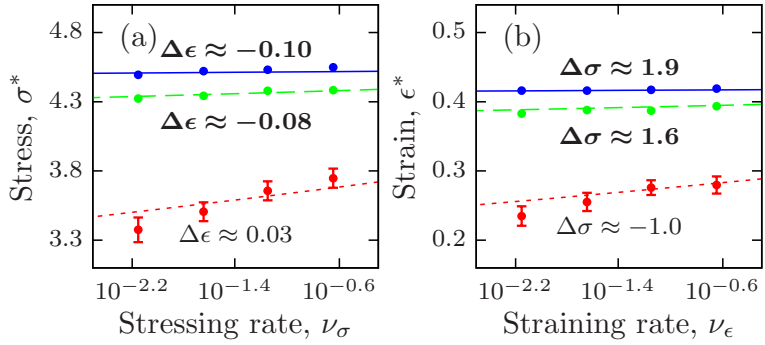}
\end{center}
\caption{Verification of our model for (a) stress- and (b) strain-controlled processes in a 2-dimensional square lattice.  Lines represent predictions of Eq.~(3) and its  $\epsilon^*$-analog,
while points and error bars (shown when bigger than the points) represent averages and standard deviations of $10$ MD simulations for a $20\times 20$ cell. The different curves correspond to $T=10^{-4}$ (solid lines), $T=10^{-3}$ (dashed lines), and $T=10^{-2}$ (dotted  lines). The strain change, $\Delta \epsilon$, and stress change, $\Delta \sigma$, during the transitions are indicated next to the lines. 
Bold face $\Delta \epsilon$ and $\Delta \sigma$ correspond to inverted responses.
 \label{fig4}}
\end{figure}

In our simulations, 
we use the potentials
\begin{align}
V_x(x) &= 
\begin{cases} 
\frac{1}{2}k_{x_0}\left(x-x_0\right)^2, & \mbox{for } x\leq{}x_1, \\
\frac{1}{2}(k_{x_1}-k_{x_0})\left(x-x_1\right)^2+\frac{1}{2}k_{x_0}\left(x-x_0\right)^2, & \mbox{for } x>{}x_1, 
\end{cases} \\
V_y(y) &= 
\begin{cases} 
\frac{1}{2}k_{y_0}\left(y-y_0\right)^2, & \mbox{for } y\leq{}y_1, \\
\frac{1}{2}(k_{y_1}-k_{y_0})\left(y-y_1\right)^2+\frac{1}{2}k_{y_0}\left(y-y_0\right)^2, & \mbox{for } y>{}y_1. 
\end{cases}
\end{align}
These potentials have continuous first derivatives (forces) that are piecewise linear.  
(We have also observed ICTs in a variety of  smooth potentials not listed here.)
The  $V_z$ potential is the Lennard-Jones potential,
\begin{align}
V_z(z) &= \zeta\left[\left(\frac{z_0}{z}\right)^{12}-2\left(\frac{z_0}{z}\right)^6\right].
\end{align}
In order to observe ICTs in our simulations, we have chosen the parameter values $k_{x_0}=11$, $k_{x_1}=2$, $x_0=2$, $x_1=2.2$, $k_{y_0}=0.5$, $k_{y_1}=20$, $y_0=2.225$, $y_1=3.3$, $\zeta=0.26$, and $z_0=0.225$.

Figure \ref{fig4} compares the predictions of the model [Eq.~(\ref{model}) and corresponding equation for $\epsilon^*$]  with MD simulations for both stress-  and strain-controlled processes.  The model quite accurately predicts the general trends  of the metastable decay over the wide range of temperatures and stressing/straining rates that we simulated.  In all cases, smaller stressing/straining rate $\nu_{\sigma,\epsilon}$ and larger temperature $T$ tend to lead to earlier decoupling transitions. Moreover, for sufficiently low temperatures (dotted  and dashed lines), these transitions are ICTs [Fig.\ \ref{fig4}(a)] and force amplification transitions  [Fig.\ \ref{fig4}(b)].   

\begin{figure*}[t!]
 \begin{center}
\includegraphics[width=14cm]{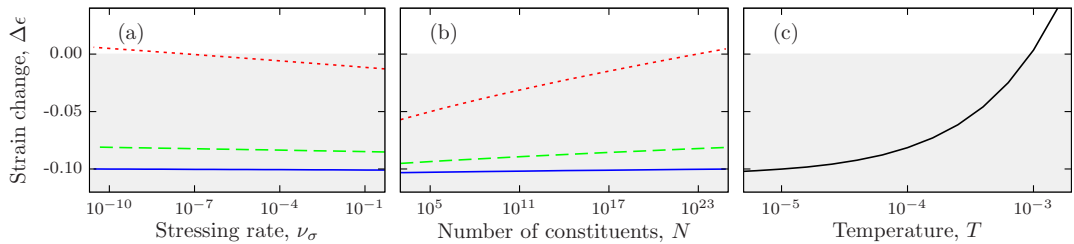}
\end{center}
\caption{Strain change, $\Delta\epsilon$, during decoupling transitions in an 
isotropically stressed 3D cubic lattice: dependence on  (a) stressing rate, (b) system size, and (c) temperature. 
In (a) and (b), the different curves correspond to $T=10^{-5}$ (solid lines), $T=10^{-4}$ (dashed lines), and $T=10^{-3}$ (dotted lines). 
When not specified, $N=10^{23}$ 
and $\nu_{\sigma}=10^{-10}$.  
The shaded regions correspond to ICTs.
Note that, because the strain is defined relative to the zero-stress size, $100\times \Delta\epsilon$ is a percentage change.
\label{fig5}}
\end{figure*}

Remarkably, as illustrated in Fig.~\ref{fig5}, our model predicts ICTs even for very large, macroscopic sizes ($N=10^{23}$) over  very small,  macroscopic stressing rates ($\nu_{\sigma}=10^{-10}$) for moderate temperatures ($T=10^{-3}$).
In fact, this persistence of ICTs can be deduced by inspection from Eqs.~(\ref{decouple}) and (\ref{model}).  
ICTs will occur provided  metastable states persist until the stress becomes sufficiently large,
while larger temperature $T$, smaller stressing rate $\nu_{\sigma}$, and larger system size $N$ all tend to promote earlier metastable decay. 
But
the temperature dependence in Eq.~(\ref{decouple}) is exponential, while the dependence on $\nu_{\sigma}$ and $N$ are only polynomial.  
Therefore,
on inverting Eq.~(\ref{model}), it follows that $\sigma^*$  depends only logarithmically on $\nu_{\sigma}$ [Fig.~\ref{fig5}(a)] and $N$ [Fig.~\ref{fig5}(b)], so that small temperatures can easily offset their effects [Fig.~\ref{fig5}(c)].  This provides evidence that ICTs are possible for a wide range of conditions.

\section{Final Remarks}
\label{concl}

It is instructive to compare phase transitions exhibiting inverted compressibility with ordinary phase transitions, which do not.  In ordinary first-order phase transitions, the curve of spinodal points in the phase diagram lies strictly within the region of phase coexistence.  The emergence of an ICT implies a violation of this condition, as follows.  For  metastable coupled states reached while increasing $\sigma$, the stress is larger than the stress of any state in the phase coexistence region (see Fig.~\ref{fig3}). 
Thus,  by convexity,   
the thermodynamic equilibrium to which  these metastable coupled states decay must have a strain larger than the strain of any state within the phase coexistence region.  Therefore, if these decays correspond to a decrease in the strain (as observed in  ICTs), the spinodal point at which the metastable coupled states cease to exist must occur at a yet larger strain, and hence the curve of spinodal points cannot be confined to the region of phase coexistence. 
Although their potential for ICTs has not been previously recognized, the global behavior of spinodal curves has been of wide interest in other contexts for many years (see, e.g., Refs.~\cite{Binder83,stanley}). Furthermore, while ICTs have not been experimentally observed,  we propose that with the significant advances made in designing both electromagnetic \cite{Smith04} and acoustic \cite{Fang06} 
metamaterials,  it is possible to realize engineered materials that undergo ICTs.

Now, one may wonder why ICTs have not been previously predicted.  This possibility was likely not anticipated because most previous continuum \cite{Falk80,Knowles93} and  lattice \cite{Truskinovsky06,Fraternali2011}  models of solid-solid phase transitions have postulated a free energy that is a function of a single dilation or shear strain field.  Such models necessarily result in  stress-strain relations in which the stress is a single-valued function of the strain, as the one shown in Fig.~\ref{fig3}(b).  However, in systems like ours, additional fields are required to model not only the total length $h$ of the constituents (corresponding to the usual strain field), but also the internal $x$ and $y$ degrees of freedom.
The many-to-many nature of the relationship in Fig.~\ref{fig3}(a) is the result of ``projecting away'' these internal strains.  More generally, this observation indicates that any effective field theory that can account for ICTs 
will necessarily require additional fields (e.g., internal strains) beyond the external strain.

\begin{figure*}[t!]
 \begin{center}
\includegraphics[width=10cm]{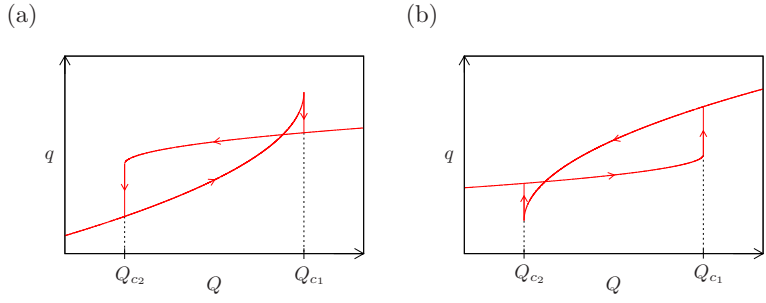}
\end{center}
\caption{
Schematics of possible  hysteresis loops of an arbitrary  thermodynamic quantity exhibiting inverted transitions.  The state variable $q$ either (a) decreases as the conjugate force $Q$ increases at $Q_{c_1}$ or (b) increases as the conjugate force $Q$ decreases at $Q_{c_2}$. These transitions occur during the decay of a metastable state.
\label{fig7}
}
\end{figure*}

Ultimately, it is the global shape of the hysteresis curve that leads to the inverted response, and thus analogous phenomena associated with thermodynamically conjugated variables other than stress and strain can also emerge.
That is, given a thermodynamic state variable $q$ (e.g., strain, entropy, magnetization, polarization, particle number) with a conjugate thermodynamic force $Q$ (e.g., stress, temperature, magnetic field, electric field, chemical potential), it generally follows from the convexity of the free energy that $\frac{\partial q}{\partial Q}>0$.  However, as depicted in Fig.~\ref{fig7}, if the system occupies a metastable state, an increase (or decrease) in $Q$ may force the system to decay to the thermodynamic equilibrium.  If this results in a decrease (or increase) in the observed value of $q$, the transition will exhibit an inverted response in which $\frac{\Delta q}{\Delta  Q}<0$ (e.g., inverted compressibility, heat-capacity, magnetic susceptibility, electric susceptibility, chemical hardness transitions). We thus suggest that the negative compressibility transitions analyzed here will have analogues in other thermodynamic properties, which may inspire the design of yet new classes of metamaterials with unusual properties.

\begin{acknowledgements}
This work was supported by an NSF Graduate Research Fellowship and the NSF Grant No.\ DMS-1057128.
\end{acknowledgements}

\section*{Appendix}

\subsection*{A.  Temperature scale $T_0$}

We define the temperature scale $T_0$ by setting $k_BT_0$ equal to the energy barrier separating the coupled and decoupled minima of the potential energy at the tension for which the values of these two minima are equal (i.e., the point at which the equilibrium phase transition would occur in the limit of small temperature).  When $T$ is much smaller than $T_0$, the energy barrier will be difficult to overcome, and metastability will persist for long periods.  
When $T$ is much larger than $T_0$, thermal fluctuations will easily overcome the energy barrier, and thermal equilibrium will prevail.

\subsection*{B. Timescale $\tau_V$}

For a $D$-dimensional cell with linear size of $n$ primitive cells,  
the most important macroscopic timescale is the inverse of the rate of the most prominent fluctuations. This corresponds to the acoustic longitudinal mode with wavenumber $\frac{2\pi}{4n}$.  For brevity, consider just a one-dimensional chain of constituents (we mention the alterations needed for the $D$-dimensional lattice at the end of the calculation).  We linearize the forces acting on the particles and consider small amplitude vibrations.  For longitudinal modes, the equations of motion are 
\begin{eqnarray}
m\frac{\partial^2u_j}{\partial t^2} &=& k_x(v_j-u_j)+k_y(w_j-u_j)-k_y(u_j-v_{j-1})-k_x(u_j-w_{j-1}),~~~ \\
m\frac{\partial^2v_j}{\partial t^2} &=& k_y(u_{j+1}-v_j)-k_x(v_j-u_j)+k_z(w_j-v_j), \\
m\frac{\partial^2w_j}{\partial t^2} &=& k_x(u_{j+1}-w_j)-k_y(w_j-u_j)-k_z(w_j-v_j),
\end{eqnarray}
where the $k$'s are the second derivatives of the corresponding potentials; $u_j$ and $u_{j+1}$ are the displacements of the exterior particles in Fig.~2(a); $v_j$ and $w_j$ are the displacements of the two interior particles.  We insert the wave ansatz $u_j=a_u\exp\left[i\left(kj-\omega t\right)\right]$, $v_j=a_v\exp\left[i\left(kj-\omega t\right)\right]$, and $w_j=a_w\exp\left[i\left(kj-\omega t\right)\right]$ to find the dispersion relation between the wavevector $k$ and the angular frequency $\omega$.  The resulting secular equation is
\begin{equation}
\det\begin{pmatrix}
-2(k_x+k_y)+m\omega^2 & k_x+k_y\exp\left(-ik\right) & k_y+k_x\exp\left(-ik\right) \\
k_x+k_y\exp\left(ik\right) & -(k_x+k_y+k_z)+m\omega^2 & k_z \\
k_y+k_x\exp\left(ik\right) & k_z & -(k_x+k_y+k_z)+m\omega^2
\end{pmatrix}=0.
\end{equation}
The acoustic longitudinal mode with wavevector $k=\frac{2\pi}{4n}$ then has period 
\begin{equation}
\label{volumefluctuation}
\tau_V=4\sqrt{\frac{3m\left(k_x+k_y+2k_z\right)}{2k_xk_y+(k_x+k_y)k_z}}n
\end{equation}
to lowest order in $1/n$.  This can be expressed as $\tau_V=4\sqrt{\frac{m_{tot}}{k_{eff}}}n$, where $m_{tot}$ is the total mass of the primitive cell and $k_{eff}$ is the effective spring constant for the primitive cell (as found from the standard expressions for series and parallel configurations).  When we consider the full lattice rather than a one-dimensional chain, the only alteration is that $k_{eff}$ is the effective spring constant for the lattice's primitive cell (which remains unchanged for cubic lattices) and $m_{tot}$ is the corresponding  mass $(1+Z)m$ of the primitive cell, where $Z$ is the coordination number. In the numerics, we use the zero-stress value of $\tau_V$ as a reference timescale.  The optical (longitudinal and transverse) modes in Eqs.\ (\ref{long})-(\ref{long2})  are derived through similar analyses.

\end{document}